\documentclass[12pt,preprint]{aastex}

\slugcomment{Accepted for publication in the ApJ Letters on 19/06/2014}

\shorttitle{First direct measurements of transverse waves in solar polar plumes using SDO/AIA }
\shortauthors{Thurgood, Morton \& McLaughlin}

\begin{document}

\title{First direct measurements of transverse waves in solar polar plumes using SDO/AIA }

\author{J.O. Thurgood, R.J. Morton and J.A. McLaughlin}
\affil{Department of Mathematics and Information Sciences, Northumbria University,
 Newcastle Upon Tyne, NE1 8ST, UK}
\email{jonathan.thurgood@northumbria.ac.uk}

\begin{abstract}
There is  intense interest in determining the precise contribution of Alfv\'enic waves propagating along solar structures to the problems of coronal heating and solar wind acceleration.
 Since the launch of SDO/AIA, it has been possible to resolve transverse oscillations in off-limb solar polar plumes and recently McIntosh et al. (2011, \nat, 475, 477) concluded that such waves are energetic enough to play a role in heating the corona and accelerating the fast solar wind. However, this result is based on comparisons to Monte Carlo simulations and confirmation via direct measurements is still outstanding. 
Thus, this letter reports on the first direct measurements of  transverse wave motions in solar polar plumes.
Over a 4 hour period, we measure the transverse displacements, periods and velocity amplitudes of 596 distinct oscillations observed in  the 171 \AA\, channel of SDO/AIA. 
We find a broad range of non-uniformly distributed parameter values which are well described by log-normal distributions with peaks at $234$ km, $121$ s and $8$ km s$^{-1}$, and mean and standard deviations of $407\pm297$ km, $173\pm118$ s and $14\pm10$ km~s$^{-1}$. 
Within standard deviations, our direct measurements are broadly consistent with previous results.
{However, accounting for the whole of our observed non-uniform parameter distribution we calculate an energy flux of $9-24$~W~m$^{-2}$, which is $4-10$ times below the energy requirement for solar wind acceleration. 
Hence, our results indicate that transverse MHD waves as resolved by SDO/AIA cannot be the dominant energy source for fast solar wind acceleration in the open-field corona.}

\end{abstract}


\section{Introduction}

The precise nature of coronal heating and solar wind acceleration remains an outstanding puzzle in solar physics, to which two broad classes of process are thought to contribute (namely, wave-based and reconnection-based, see e.g. Erd{\'e}lyi \& Ballai \citeyear{2007AN....328..726E}, Cranmer  \citeyear{2009LRSP....6....3C}, \citeyear{2012SSRv..172..145C}, Hansteen \& Velli \citeyear{2012SSRv..172...89H}). In the wave-driven models, it is typically proposed that propagating Alfv\'enic waves 
 are responsible for the transport of magnetoconvective energy upward into the solar corona, guided along regions of predominantly open magnetic flux, such as polar coronal holes. This mechanical energy is then subsequently deposited as thermal energy
 via various ancillary dissipation mechanisms such as phase mixing (Heyvaerts \& Priest \citeyear{HP83}), resonant absorption (e.g. Goossens et al. \citeyear{Goossens2011}), wave interactions and turbulent cascade (e.g. Matthaeus et al. \citeyear{1999ApJ...523L..93M}), and nonlinear compressibility, steepening and shock formation  (e.g. Ofman \& Davila \citeyear{1997ApJ...476..357O}, Suzuki \& Inutsuka \citeyear{2005ApJ...632L..49S}).
Note that in this context the term Alfv\'enic wave is used in the broad sense of a transverse, magnetic-tension-driven magnetohydrodynamic (MHD) wave; the specific identity of such waves is dependent on the structuring of the wave-guide (see e.g., Erd{\'e}lyi \& Fedun \citeyear{2007Sci...318.1572E}, Van Doorsselaere et al. \citeyear{2008ApJ...676L..73V}, Goossens et al. \citeyear{2009A&A...503..213G}).

Determining the prevalence of transverse MHD waves and their energetic properties in the open-field corona is therefore of particular importance with regards to validating heating models and assessing their contribution compared to other mechanisms. To date, the majority of quantifiable evidence for transverse MHD waves propagating in coronal holes is indirect, obtained via measuring the non-thermal broadening of spectral lines (see Banerjee et al. \citeyear{2011SSRv..158..267B} for a review). Despite the challenges in separating observed line-broadening into thermal and non-thermal contributions (Dolla \& Solomon \citeyear{2008A&A...483..271D}) many studies  have shown non-thermal velocities ($v_{\mathrm{nt}}$) that are consistent with the presence of Alfv\'enic waves and have reported large ranges of $v_{\mathrm{nt}}$, which vary with altitude (e.g. Doschek \& Feldman reported $v_{\mathrm{nt}}\approx20$ km s$^{-1}$ at 30\arcsec\, above the limb, 
whereas Hahn \& Savin \citeyear{2013ApJ...776...78H} find $v_{\mathrm{nt}}\approx 40$ km s $^{-1}$ at around 190\arcsec).   The reported non-thermal velocities have also been found to vary between plume and inter-plume plasma by Banerjee et al. (\citeyear{2009A&A...501L..15B}), where higher non-thermal velocities are reported in the inter-plume regions.
Spectroscopic studies have also shown that with increasing altitude, beyond around $1.12R_{\odot}$, these non-thermal velocities begin to fall short of the trend that is expected for undamped Alfv\'enic waves ($\propto n_{e}^{{-1}/{4}}$) and this has been interpreted as evidence of Alfv\'enic wave damping at sufficiently low altitudes to permit significant contributions to heating and acceleration (Bemporad \& Abbo \citeyear{2012ApJ...751..110B}, Hahn et al. \citeyear{2012ApJ...753...36H}, Hahn \& Savin \citeyear{2013ApJ...776...78H}).

Despite this abundance of spectroscopic evidence, direct detection and measurement of transverse waves propagating in coronal holes via imaging  is desirable because other non-oscillatory mass motions may also contribute to $v_{\mathrm{nt}}$. Although transverse waves have been found to be ubiquitous elsewhere in the Sun, 
such as in chromospheric structures (see, e.g. De Pontieu et al. \citeyear{2007Sci...318.1574D}, Morton et al. \citeyear{Morton2012}) prominences (e.g. Hillier et al. \citeyear{2013ApJ...779L..16H}) and in coronal loops (see, e.g. Aschwanden et al. \citeyear{1999ApJ...520..880A}, Aschwanden \& Schrijver \citeyear{2011ApJ...736..102A}, White \& Verwichte \citeyear{2012A&A...537A..49W}) 
so far only one paper has considered the presence of transverse waves in the faint, plume-like structures visible in EUV above coronal holes.    McIntosh et al. (\citeyear{McIntosh2011}) demonstrated that signatures of transverse wave motion could be found in off-limb structures using  data from \textit{Solar Dynamics Observatory's Atmospheric Imaging Assembly}  (SDO/AIA data) and they determined that the observational data was visually compatible with synthetic data created by Monte Carlo simulation of oscillating features with velocity amplitudes of  25$\pm$5 km s$^{-1}$ uniformly distributed across a period range 150-550 s. 
{These parameters imply that the observed waves carry sufficient energy flux to facilitate solar wind acceleration (models require $100-200$ W~m$^{-2}$ of wave energy flux near the coronal base to match empirical data, see, e.g., Withbroe \citeyear{1988ApJ...325..442W}, Hansteen \& Leer \citeyear{1995JGR...10021577H}).}
However, it is important to note that  whilst comparison to a Monte Carlo simulation can be informative (acting to constrain key parameters), direct measurements of the amplitudes and periodicities of these oscillations are still a necessity. 

This letter addresses this need and  presents the first direct measurements of transverse wave motion in the faint, off-limb plume-like features that can be resolved by the SDO/AIA. In this letter, we present measurements of 595 transverse oscillations seen in the 171 \AA\, channel of AIA 
at 5 different altitudes above the solar limb during a 4 hour period. We detail the observations and methods used in Section \ref{methods}, present our results in Section \ref{results} and briefly discuss their implications in Section \ref{discussion}.

\section{Observations and Analysis}\label{methods}

\subsection{Observations}
\begin{figure}
\epsscale{.80}
\plotone{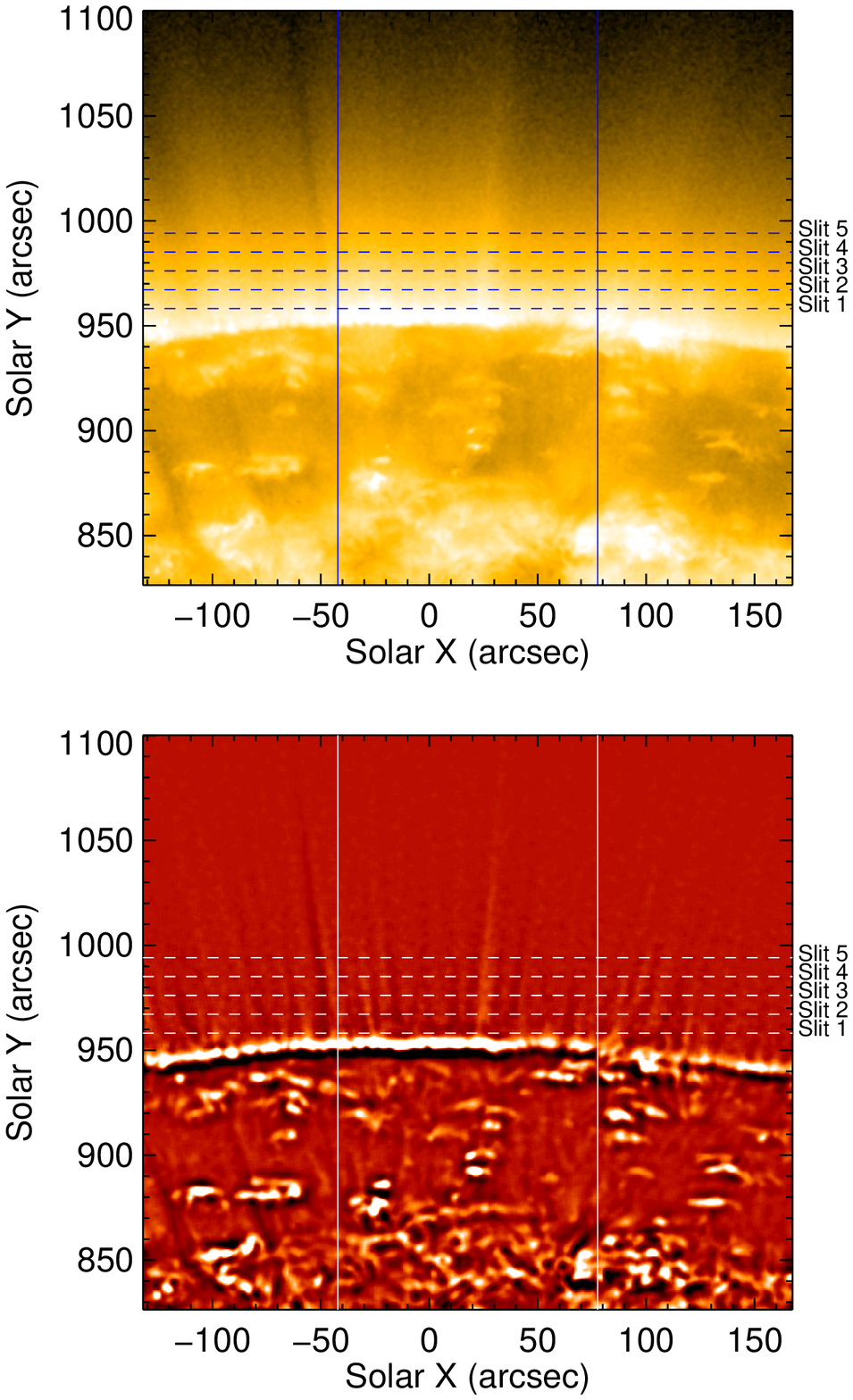}
\caption{
The Solar north pole as seen by SDO/AIA in 171 \AA\, on the 6 August 2010 at 00:00 UT. The top panel shows Level 1 data on a log-scale, and the bottom shows the enhanced, unsharp-masked image on a linear scale. The dashed lines show the altitudes of 5 synthetic slits used for the data analysis (see Table \ref{table}), which are 200 pixels ($\approx$ 87 Mm) long and run between the solid lines. A movie of the bottom panel is available in the online edition.
\label{referencemap}}
\end{figure}

We consider SDO/AIA data taken in the 171\AA$\,$ channel around the northern pole on  6$^{\mathrm{th}}$ August 2010 between 00:00 UT and 04:00 UT (Figure \ref{referencemap}). The data has pixel size of 0.59\arcsec\,, a cadence of 12 s, and a 2.7 s exposure time. 
The alignment of the data was tested using the IDL routine  \textit{FG\_RIGIDALIGN.PRO} which  showed that frame-to-frame jitter was consistently smaller than the displacement accuracy of the alignment routine (this was found to be the case for all coalignment windows tested). Thus no alignment is required, although  we do account for sub-pixel jitter as a source of uncertainty in our analysis (Section \ref{methods2}). No derotation is performed, as rotation near the poles is negligible compared to the feature lifetimes.
To better reveal the off-limb features, the data is subject to an unsharp masking procedure, in which each image in the sequence has a $9\arcsec\times 9\arcsec$ boxcar-smoothed image removed from itself. This preserves the finer features whilst removing those at larger scales. The resulting sequence is then further smoothed over 3 time steps, which suppresses frame-to-frame variations in intensity and aids the fitting routine described in section \ref{methods2}. In the enhanced data, we clearly see off-limb plume structures which often exhibit transverse motion by eye (Figure \ref{referencemap} and supplementary movie).

\subsection{Analysis}\label{methods2}

\begin{figure}
\plotone{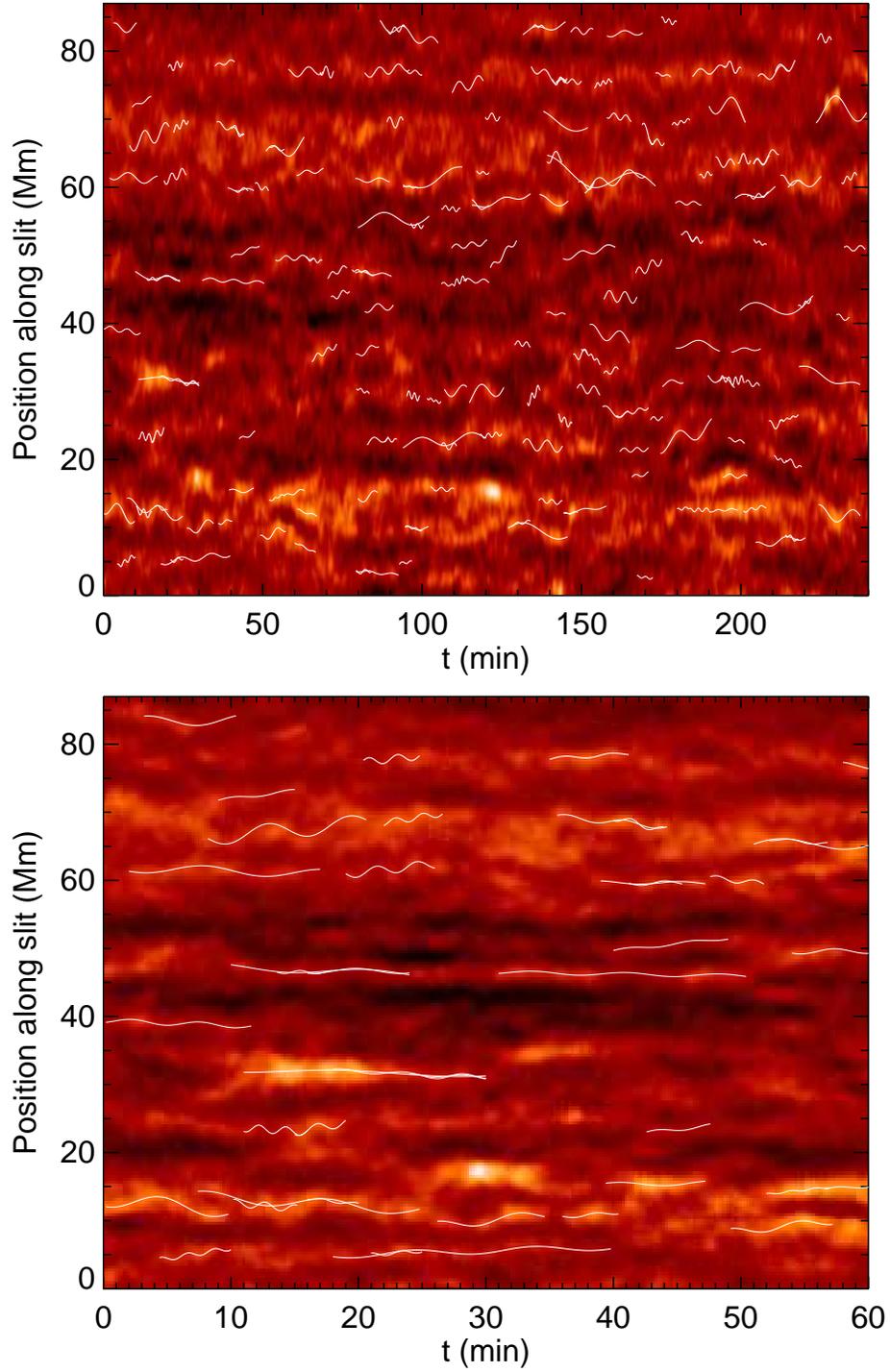}
\caption{Time distance map from Slit 1 for the whole time series (top panel) and the first hour (bottom panel). A selection (i.e., not full sample) of longer-period fits that were made to the oscillating features are  overlaid in white.}\label{timedistancemap}
\end{figure}

We measure transverse oscillations of these structures by considering time-series data from 5 synthetic slits, which are 200-pixels long and placed at increasing altitudes above the limb (see Figure \ref{referencemap} for positions and Table \ref{table} for altitude information). To improve S/N, the intensity at each position is taken as the mean of the slit intensity across the 5 pixels (i.e. artificial slits of $1\times200$ pixels are created from $5\times200$ pixel data cut-outs).
An example of the resulting  time-distance diagrams is shown in Figure \ref{timedistancemap}. They are characterised by intermittent bright streaks (corresponding to the location of the over-dense, plume structures) which, by eye, exhibit clear signs of transverse motion, many of which appear to oscillate sinusoidally for one or more cycles. On visual inspection it is clear that these oscillations are not all similar, in the sense that it appears as if there is a distribution of oscillations with different periods and displacement amplitudes, and this is further complicated by variable feature lifetimes.

To detect the central position of the plume axes with sub-pixel accuracy, and follow their evolution through the time series, we employ an automated Gaussian-fitting method developed by  Morton et al. (\citeyear{2013ApJ...768...17M}) and Morton \& McLaughlin (\citeyear{2013A&A...553L..10M}). First, for each time, local intensity peaks in X (position along slit) are located for a 10-pixel neighbourhood, whereby a crawling algorithm checks all neighbourhoods $0\le X\le 10,1\le X\le 11, ..., 190\le X\le200$. Then, the gradients either side of the determined location are checked against a threshold gradient of 0.5. If the gradient is sufficient, the point is considered a local maxima. Note that the threshold gradient is determined experimentally, by visually comparing the pixel-locations of maximum intensity against the time distance maps. Once pixels containing local intensity maxima are determined, their position is then refined to sub-pixel accuracy by fitting the 5-pixel neighbourhood  (centered around the point) with a Gaussian. The fitting is weighted by AIA 171 \AA\, intensity errors which are taken as 
\begin{equation}
\sigma_{\mathrm{noise}}(F) \approx \sqrt{2.3 +0.06F} / \sqrt{5} \quad(\mathrm{DN})
\end{equation}
as per Yuan \& Nakariakov (\citeyear{2012A&A...543A...9Y}), where $F$ is the pixel-intensity of the unaltered, Level 1 data and a product of $1/\sqrt{5}$ is taken because the time-distance diagrams are constructed from the average of 5-neighbouring slices.
The uncertainty on the position of the local maxima is taken as the $\sigma$ estimate on the position  of the apex of the fitted  Gaussian  and the uncertainty on position due to sub-pixel jitter, added in quadrature. The jitter is calculated as the root mean square of the frame-to-frame displacements calculated by \textit{FG\_RIGIDALIGN.PRO}, which gives $\sigma_{\mathrm{jitter}} \approx 0.03$ pixels.  These local intensity maxima are then traced through the time series.
Finally, any followed threads which have less than 20 points (persist for less than 240 s) are rejected.

On inspection, these automatically-detected threads show clear signs of oscillatory behaviour as per our earlier visual impression from the time-distance diagrams. Using  Levenberg-Marquardt least-squares minimization (Markwardt \citeyear{2009ASPC..411..251M}) we fit the data with a sinusoid and linear trend, i.e. a function of the form
\begin{equation}
X[t]=\xi\sin\left(\frac{2\pi t}{P} + \varphi \right) + At + B \,,
\end{equation}
from which we determine the {maximum} transverse displacement $\xi$, the period $P$, and thus calculate the {maxiumum} velocity amplitude $v=2\pi \xi P^{-1}$.
Because the tracking-routine picks up threads of variable lengths (ranging from $20$ to $450$ time-steps or \lq{points}\rq{}, due to variable feature lifetimes), longer threads can often be fitted with different oscillations during different stages of their lifetime, and it is often the case that shorter-period oscillations can be seen super-imposed upon longer period trends. Thus, where multiple fits to different subsections of a longer thread are possible, all such fits are taken and contribute to the sample, provided they meet the following selection criteria (which applies to all fits made):
\begin{itemize}
\item{At least $3/4$ of an oscillatory cycle must be observed.}
\item{The fit must be made using more than 5 points (this effectively renders the minimum period of oscillation around 60 s).}
\item{Fits dependent on a set of points that includes a jump of more than 2.5-pixels from point-to-point are discounted, as this constitutes an instantaneous transverse velocity of greater than 100 km s$^{-1}$ which is an order of magnitude greater than the representative velocity amplitudes measured for all samples (even when this criteria is relaxed). There is no restriction on maximum transverse displacements measured, but transverse motion must be well-resolved, otherwise it is thought that such jumps are artefacts from the thread-following algorithm.}
\item{Errors on fitted parameters must not be comparable in magnitude to the fitted parameter itself. In practice, we reject a fit if the fractional error on velocity amplitude $\sqrt{\delta\xi^{2}+\delta P^{2} }>0.7$.}
\end{itemize}

\begin{figure}
\plotone{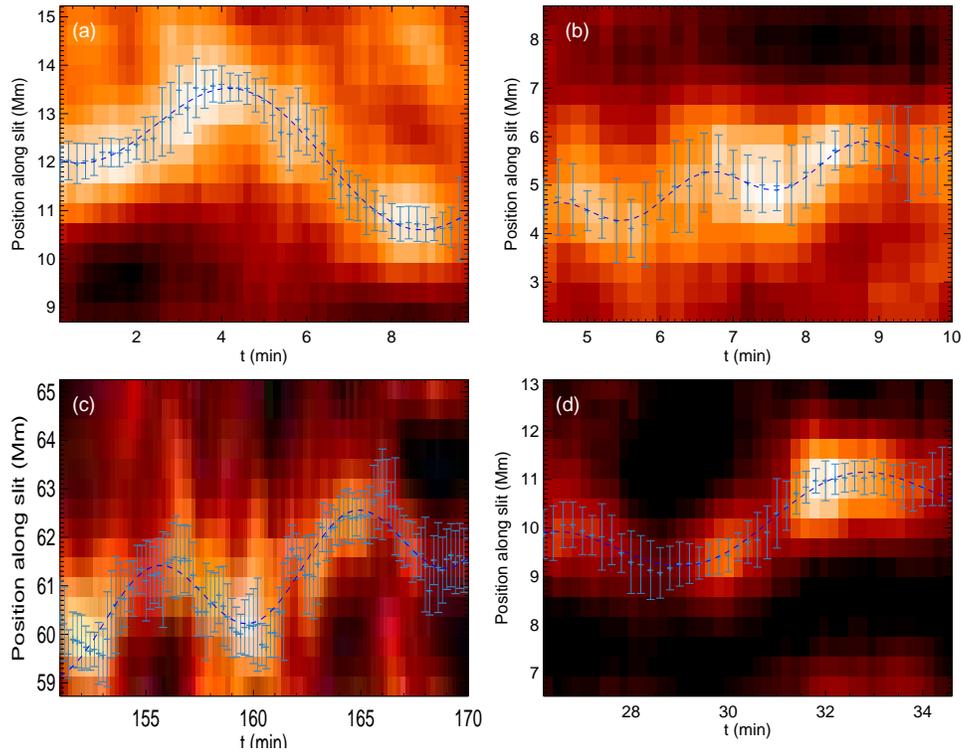}
\caption{Close-up view of fitted oscillations. The center of the blue vertical error bars show the feature center from time-to-time as determined by the Gaussian fitting routine  and the $\pm \sigma$ uncertainty on that position. The blue-dashed curve through the feature shows the best sinusoidal fit to the feature center, from which the wave parameters are derived.
\label{example_fits}}
\end{figure}

The measurements made for Slit 1 are shown in Figure \ref{timedistancemap} and, as a typical example, 4 fits are shown in further detail in Figure \ref{example_fits}.

\section{Results}\label{results}
The detection and fitting procedure detailed in Section \ref{methods} is applied to 5 slits placed at different altitudes above the limb. Information regarding slit position, number of threads detected and number of fitted oscillations is summarised 
in Table \ref{table}. At higher altitudes, fewer threads are detected and followed throughout the time-distance diagram. This is because at higher altitudes many of the smaller plumes become increasingly diffuse and thus have poorer contrast against the backgrounds of the time-distance diagrams (cf. Figure \ref{referencemap} and column 3 of Table \ref{table}).

For all 5 slits, we find a broad distribution of oscillations with transverse displacements ranging from of $64$ to $2558$ km, periods  of $61$ to $2097$ s, and velocity amplitudes of $1$ to $88$ km s$^{-1}$. 
The mean values for transverse displacement $\xi$, period $P$ and velocity amplitude $v$ are given for each slit in Table \ref{table}. Within the standard deviations, there is no appreciable change in mean parameters at different altitudes. Given this, if we pool all 596 measurements we find a mean displacement of $424\pm277$ km,  mean periods  $233\pm207$ s, and mean velocity amplitudes $14\pm8$ km s$^{-1}$. 
However, due to the relatively broad range of measured parameters, these simple averages may be sensitive to outliers and may not be completely representative of the distribution of measurements. 

\begin{figure}
\epsscale{.55}
\plotone{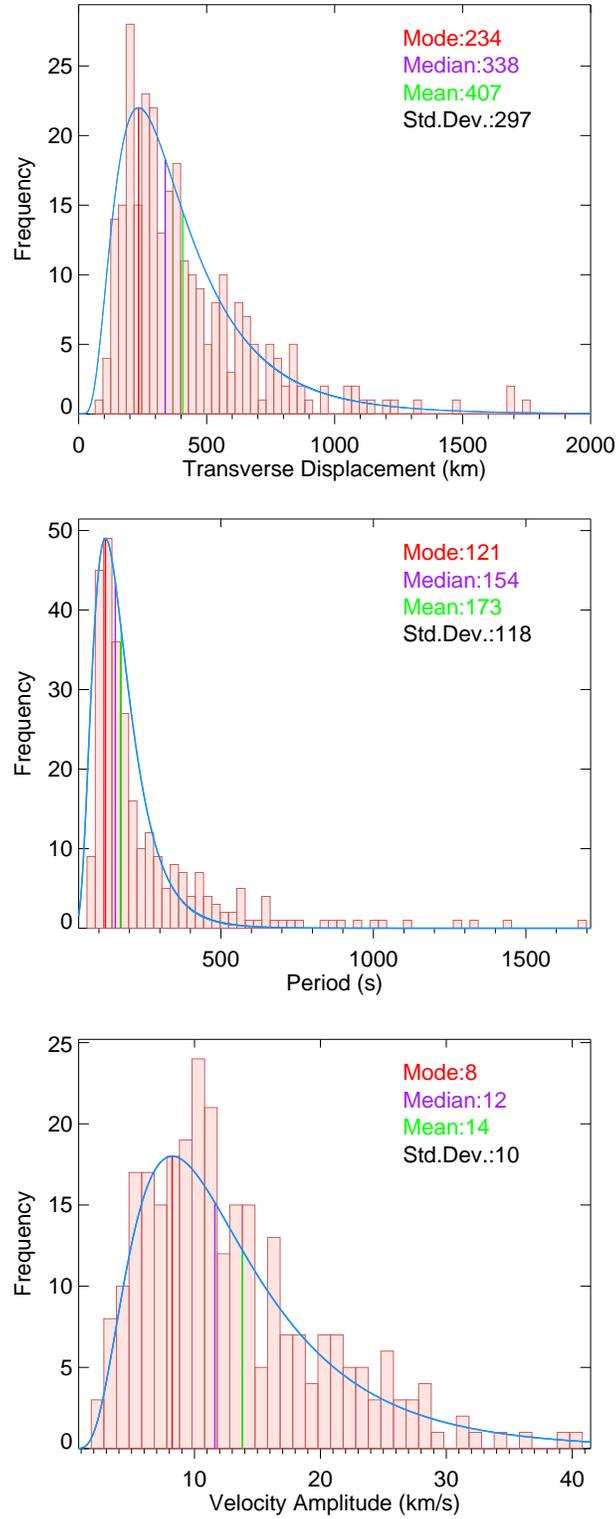}
\caption{Histograms showing the parameter distributions from Slit 1. The distributions show a clear positive skew, and can be fitted with a log-normal distribution (blue curve). The mode, median, mean and standard deviations according to the log-normal distribution are shown.}
\label{distro}
\end{figure}

Upon constructing histograms for each slit, we find that the distribution of all three parameters typically has a positive skew, with long period, large displacement and large amplitude measurements occurring infrequently in the tail (see Figure \ref{distro}).
 We find that a Logarithmic-Normal distribution fits the observed distribution well, from which we can infer an alternative quantitative description of the measured parameters. 
From Figure \ref{distro} we see that the distribution for Slit 1 peaks at a transverse displacement of $234$ km, period of $121$ s and velocity amplitude of $8$ km s$^{-1}$ (modes) and has (log-normal) mean $\pm$ standard deviation parameters of $407\pm297$ km, $173\pm118$ s and $14\pm10$ km s$^{-1}$. This log-normal mean can be thought of as a weighted mean according to the fitted probability distribution function.
This distribution (and Figure \ref{distro}) is also representative of Slits 2 and 3, qualitatively in that the data is positively skewed, and quantitatively in that the mode, median and mean of the parameters are similar (see Table \ref{table}). However, histograms for Slits 4 and 5 prove to be too underpopulated to convincingly fit any distribution function. 
Nonetheless, for all slits it is clear that transverse oscillations are detected with different displacements, periods and velocity amplitudes that occur in a fairly broad  range, but that, at least for Slits 1-3, displacements of around  $234$ km, periods of around $121$ s, and velocity amplitudes of around  $8$ km s$^{-1}$ are most commonly observed.

\section{Discussion}\label{discussion}
In this letter, we find that transverse MHD waves propagate in solar off-limb polar plumes with parameter values that are non-uniformly distributed over a broad range, with transverse displacements 
from $64$ to $2558$ km, periods  from $61$ to $2097$ s, and transverse velocities from $1$ to $88$ km s$^{-1}$. 
 We also detect multiple, superimposed oscillations of different amplitudes and periods. 
This is likely due to the fact that the oscillations are excited in the lower solar atmosphere as wave packets of finite duration (similar behaviour has also been seen for other magneto-acoustic waves, e.g. McIntosh \& Smillie \citeyear{2004ApJ...604..924M}).
The measured parameter distributions have a significant positive skew and are well described by a log-normal distributions.

Our findings for arithmetic and log-normal mean parameters are generally smaller than (but consistent within their standard deviations)  with previous results. For example, the Monte Carlo simulations of McIntosh et al. (\citeyear{McIntosh2011}) indicated that periods of $150-550$ s and velocity amplitudes of $25\pm5$ km s$^{-1}$ were comparable to the typical features of the time-distance diagrams. 
Our observed transverse velocity amplitudes are also broadly compatible within standard deviations with the non-thermal velocities reported in plume plasma by Banerjee et al. (\citeyear{2009A&A...501L..15B}), who reported $v_{\mathrm{nt}}=22.1$ km s$^{-1}$ and $v_{\mathrm{nt}}=25.9$ km s$^{-1}$ (for 195 \AA\, and 202 \AA\, lines of EIS respectively)  at an altitude of around $9$ Mm. 
Assuming that this non-thermal velocity measurement contains contributions from both transverse and torsional MHD waves,
this comparability suggests that torsional Alfv\'en waves may not strongly contribute to plume non-thermal velocity measurements (otherwise we would expect $v_{nt}$ to be larger). However, it is important to note that whether this is due to a lack of larger amplitude torsional motions in plumes or simply that they are present but their line-of-sight motions {are under-resolved by current spectrometers} is currently unclear.

This letter now confirms by \textit{direct measurement} that transverse waves with such velocity amplitudes and periodicities can occur at altitudes of around $\approx8-35$ Mm, however, we have found a broader range of parameters and thus the full picture is much richer. The technique used for the direct measurement of imaging data also overcomes the inability of the Monte Carlo method to reveal information on the nature of oscillations with displacements on the order of or smaller than the diffraction limit of SDO. 
It has been said that these waves carry enough energy flux to make a significant energy contribution; as an estimation McIntosh et al. (\citeyear{McIntosh2011}) calculate an energy flux using the formula $E_{A}=\rho v_{A} f v^{2}$ (an equation for Alfv\'en waves in a homogeneous plasma, where coronal density is taken as  $\rho\approx\left[5-10\right]\times10^{-13}$ kg m$^{-13}$ and phase-speed as $v_{A}\approx 200-250$ km s$^{-1}$). By assuming a filling factor of $f\approx 1$, {they find} $E_{A}=100-200$~W~m$^{-2}${, which is compatible with  the wave energy requirement for solar wind acceleration}. 
 However, because we have found a broad distribution of parameters which is skewed in favour of relatively smaller amplitude waves, we can now take the whole range of parameters into account when calculating $E_A$.

Appropriate weighting can be given to the range of velocity amplitudes by taking the volume filling velocity $v$ to be the log-normal mean (i.e., 
$\int Fv\,dv$ where $F$ is the log-normal probability density function; evaluated over the whole range). This is equivalent to a weighted average. For Slit 1 we find $E_{A}=19-47$ W m$^{-2}$; a quarter of the previous estimate. Furthermore, this calculation {is an overestimation in that it} does not provide the time averaged energy flux. {Replacing the (maximum) velocity term with its mean value over a period introduces a factor of $1/2$},  i.e.~$\langle v(t) \rangle =v /\sqrt{2}$, giving $E_{A}=9-24$ W m$^{-2}$ which indicates that the transverse waves transport a less significant amount of energy through the corona.

{
This energy budget, determined by direct measurement, falls  $4-10$ times below the minimum theoretical requirement. Furthermore, it is calculated according to the equation for volume-filling Alfv\'en waves in a homogeneous plasma, which corresponds to the most generous case in terms of energy flux. More realistic assumptions about the structuring and inhomogeneity of the medium would yield smaller energy fluxes still, as discussed by Goossens et al. (\citeyear{2013ApJ...768..191G}). Thus, \emph{the evidence suggests that transverse waves in polar coronal holes are insufficiently energetic to be the dominant energy source of the fast solar wind}. Crucially, our measurements show a deficit in the energy budget independently of the debate about the most appropriate theoretical model for these waves. Contrary to the conclusions of McIntosh et al. (\citeyear{McIntosh2011}), unless there is a prevalence of large amplitude, short period transverse oscillations below the cadence of AIA, transverse waves cannot be the dominant energy source of the fast solar wind in the open-field corona and so we must now consider the contributions of alternative energy sources in coronal holes, whether it be other wave modes (such as the aforementioned torsional Alfv\'en waves) or other non-wave processes.
}

Whether our observed distributions are truly physically representative of transverse wave motion in solar polar plumes is an important question. For instance, it is likely that many shorter period oscillations are beyond the resolution of SDO/AIA  and that very large period oscillations are obscured by short feature lifetimes limiting observations of the lowest-frequency waves. Additionally, due to the presence of multiple, superimposed oscillations during a given feature's lifetime, there is a possibility of over-sampling of the most  visible parameter ranges. In this letter, for longer lived plumes we have often been able to make one measurement of an oscillation of the order of its lifetime, and several measurements of faster oscillations. Because we are measuring the signals of \textit{propagating waves} (packets as opposed to standing waves) we have considered every oscillation that meets the fitting criteria in our sample and believe that the prevalence of relatively smaller parameter ranges is not due to oversampling or selection bias. Regardless, the above calculation for $E_{A}$ illustrates that when assessing  wave energy flux, it is important to consider the broadband, ensemble behaviour of the waves and their relevant prevalences. Our observed distribution of parameters provides a point of comparison for future models of transverse waves propagating in \lq{realistic}\rq{} coronal holes (in the geometric/structuring sense of variable flux tube strengths, merging heights, canopy heights and driving motions throughout the hole), the need for which has been discussed by authors such as Cranmer \& Van Ballegooijen (\citeyear{2005ApJS..156..265C}, their section 8).

\acknowledgments

\textit{Acknowledgements:} Effort sponsored by the Air Force Office of Scientific Research, Air Force Material Command, USAF, under grant number FA8655-13-1-3067. The U.S Government is authorized to reproduce and distribute reprints for Governmental purpose notwithstanding any copyright notation thereon. The authors acknowledge IDL support provided by STFC.  R.J.M. is grateful to Northumbria University for the award of the Anniversary Fellowship.


\clearpage


\begin{deluxetable}{ccccccccccccc}
\tabletypesize{\scriptsize}
\rotate
\tablecaption{ Slit information and key parameters \label{table}}
\tablecolumns{13}
\tablewidth{0pc}
\tablehead{
~&~&~&~&  \multicolumn{3}{c}{ARITHMETIC MEAN $\pm$ 1 STD.DEV.} &\multicolumn{3}{c}{LOG-NORMAL MEAN $\pm$ 1 STD.DEV.}&\multicolumn{3}{c}{LOG-NORMAL MODES}
\\
\colhead{Slit} & \colhead{Altitude} & \colhead{No. of} & \colhead{No. of } & \colhead{ Displacement} 
& \colhead{Period} & \colhead{Velocity Amplitude} & \colhead{ Displacement} 
& \colhead{Period} & \colhead{Velocity Amplitude}& \colhead{ Displacement} 
& \colhead{Period} & \colhead{Velocity Amplitude}\\
 No. & (Mm)& \colhead{Threads} & \colhead{Fits} & \colhead{(km)} 
& \colhead{(s)} & \colhead{(km/s)} & \colhead{(km)} 
& \colhead{(s)} & \colhead{(km/s)} & \colhead{(km)} 
& \colhead{(s)} & \colhead{(km/s)} 
}
\startdata
    1    &          8.71                  &  		189                        &  		281                           &  		429$\pm$
 		293     &  		258$\pm$
 		232   &  		13$\pm$
 		7  & 407$\pm$297 & 173$\pm$118 & 14$\pm$10 &	 		
		234 & 121 & 8 		
 		\\
    2    &        15.23                    &  		159                        &  		132                           &  		498$\pm$
 		314   &  		220$\pm$
 		136   &  		16$\pm$ 		10  &
 		498$\pm$349 & 200$\pm$141 & 17$\pm$12 &
 		323 & 126 & 11
 		\\
    3    &             21.74               &  		105                        &  		99                            &  		360$\pm$
 		205     &  		220$\pm$
 		247   &  		14$\pm$ 		8 &
		 353$\pm$246 & 156$\pm$106 & 14$\pm$10 &		
		 235 & 114 & 9
 		\\
    4    &                  28.26          &  		54                         &  		51                            &  		386$\pm$
 		220     &  		189$\pm$
 		180   &  		16$\pm$ 		10 &
 		-&-&-&-&-&-\\
    5    &                       34.77     &  		26                         &  		33                            &  		338$\pm$
 		143     &  		174$\pm$
 		64 &  		14$\pm$
 		8 &
 		-&-&-&-&-&-\\
\enddata
\end{deluxetable}

\end{document}